\documentclass[aps,prb,amsmath,twocolumn,amssymb,titlepage,superscriptaddress,shownopacs]{revtex4-2}
\usepackage[english]{babel}
\usepackage{graphicx} 
\usepackage{transparent}
\usepackage{amsmath}
\usepackage{amssymb}
\usepackage{epstopdf}
\usepackage{amsfonts}
\usepackage{bm}
\usepackage{times}
\usepackage{color}
\usepackage{mathtools}
\usepackage{hyperref}
\usepackage{tabularx}
\usepackage{hhline}
\usepackage{chngcntr}

\newcommand{\astcycl}{\mathrlap{\kern0.085em{\circlearrowright}}\ast}
\newcommand{\taucycl}{\mathrlap{\kern0.42em{\bullet}}\circlearrowright}

\begin{document}

\title{Floquet X-Ray Scattering as a Probe of Hidden Electronic Orders}
\author{Martin Eckstein}
\affiliation{I. Institute for Theoretical Physics, University of Hamburg, Notkestraße 9-11, 22607 Hamburg, Germany}
\affiliation{The Hamburg Centre for Ultrafast Imaging, Hamburg, Germany}
\author{Eva Paprotzki}
\affiliation{I. Institute for Theoretical Physics, University of Hamburg, Notkestraße 9-11, 22607 Hamburg, Germany}
\affiliation{The Hamburg Centre for Ultrafast Imaging, Hamburg, Germany}


\begin{abstract}
We develop a theoretical framework for Floquet resonant X-ray scattering, using Floquet theory combined with the ultrashort  core-hole lifetime expansion. We obtain a compact expression for the Floquet  components of the resonant inelastic X-ray scattering operator, which shows that  Floquet X-ray scattering provides direct access to bond and current correlations that do not directly produce charge Bragg peaks in conventional diffraction. Applying this framework to charge-ordered  states on the Kagome lattice, we demonstrate that different symmetry-breaking  orders exhibit distinct polarization fingerprints in the Floquet Bragg peaks.  Moreover, the relative weight of bond and current contributions can be tuned through the drive frequency.  These results establish Floquet X-ray scattering as a symmetry-resolved probe of hidden electronic order or fluctuations in quantum materials.
\end{abstract}

\maketitle

{\em Introduction ---}
Time-resolved X-ray scattering provides a direct probe of electronic correlations and collective excitations in nonequilibrium quantum materials, combining \AA-scale spatial resolution with elemental selectivity.  X-ray diffraction (XRD) measures density correlations in momentum space, while resonant inelastic X-ray scattering (RIXS) probes collective excitations like magnons and orbitons~\cite{Ament2011,Mitrano2024}.  The advent of free-electron lasers has enabled ultrashort X-ray pulses with sufficient temporal and energy resolution to access light-induced states of matter~\cite{Giannetti2016, Sentef2021, Murakami2025}, establishing time-resolved X-ray scattering and spectroscopy as a versatile tool to study driven quantum materials~\cite{Mitrano2020, Lojewski2024, Merzoni2025, Padma2025}.  In particular, time-resolved RIXS has been proposed as a probe of Floquet states in periodically driven systems, where the laser driving modifies band structures and effective interactions~\cite{Wang2021, Wang2020, Werner2021RIXS, Chen2020, Kim2025, MuellerGrandi2023}.  However, strong laser fields affect not only the initial and final states, but also the scattering process itself: nonlinear wave mixing between the X-rays and the drive \cite{Glover2012}  can generate Floquet sidebands, in which the scattered photon exchanges integer multiples $n\Omega$ of the drive frequency.  Laser-dressed X-ray scattering \cite{Popova-Gorelova2018,Popova-Gorelova2024Perspective} has already revealed light-induced charge-density oscillations in solids~\cite{Ornelas-Skarin2026}, and was proposed as a probe of laser-driven currents~\cite{Popova-Gorelova2015, Popova-Gorelova2024}, and excitons \cite{Sturm2026}.

Electronic orders involving bond and current degrees of freedom have been proposed in a range of quantum materials, including loop-current order in the pseudogap phase of cuprates~\cite{Varma1997,Simon2002,Varma2006} and bond-ordered states in Kagome metals~\cite{DiSante2026, Neupert2022, Fu2024, Wagner2023, Park2021, Christensen2021, Tan2021, Denner2021}. These orders often produce little  charge-density modulation and are therefore difficult to detect with conventional diffraction.  Existing experimental signatures rely on indirect probes of symmetry breaking, such as Kerr rotation and nonlinear optics.  This calls for scattering techniques that are directly sensitive to bond and current order parameters at finite momentum.  Here we focus on resonant X-ray scattering, where the core-excited intermediate state makes the process sensitive to the laser driven valence-electron dynamics.  We develop a minimal theory of resonant laser-dressed X-ray scattering applicable to both Floquet XRD (F-XRD), i.e., Bragg scattering with energy transfer $n\Omega$, and Floquet RIXS.  We show that drive-induced Floquet sidebands encode bond and current correlations at finite momentum, enabling direct detection of the corresponding electronic order parameters.

{\em General framework ---}
The general expression for time-resolved RIXS  can be written in the form~\cite{Chen2019,Eckstein2021RIXS}
\begin{align} 
	I
	=& 
	\int dt dt'  \, s(t) s(t')  
	\,e^{i\omega_{\rm l}(t'-t) }
	\,\,\,\times
	\nonumber \\
	&\times\,\,\,\langle  U(t_0,t') R_{\mathbf q}^\dagger(t') U(t',t)  R_{\mathbf q}(t)U(t,t_0) \rangle,
	\label{eq:rixs_formulasfga}
\end{align}
where $\mathbf q$ and $\omega_{\rm l}=\omega_{\rm in}-\omega_{\rm o}$ denote momentum and energy transfer, between ingoing and outgoing photon,  $s(t)$ is the probe envelope, and $U(t,t')$ the time evolution operator  (see Appendix for details of the derivations below). The scattering operator $R_{\mathbf q}$, whose form will be given later, encodes the intermediate-state dynamics of the RIXS process. For periodically driven systems, the evolution operator can be written in Floquet form
\begin{align}
U(t,t') = e^{-iK(t)} e^{-i(t-t')H_F} e^{iK(t')},
\end{align}
with the Floquet Hamiltonian $H_F$ and the time-periodic kick operator $K(t)$ \cite{Bukov2015}. This allows to define a dressed, time-periodic scattering operator
\begin{align}
R^F_{\mathbf q}(t)=e^{iK(t)}R_{\mathbf q}(t)e^{-iK(t)}
= \sum_n R^F_{\mathbf q,n} e^{-in\Omega t}.
\label{gfwehjaa}
\end{align}
With this representation, Eq.~\eqref{eq:rixs_formulasfga} can be linked to different types of measurements:  (i) Cycle-averaged Floquet RIXS:  When the system is in a time-periodic state, and the probe envelope $s(t)$ encloses many cycles, Eq.~\eqref{eq:rixs_formulasfga} reduces to a superposition of Floquet sidebands, $I_{\mathbf q}(\omega_{\rm l}) \propto
\sum_n  I_{\mathbf q,n}(\omega_{\rm l})$, where 
\begin{align}
I_{\mathbf q,n}(\omega_{\rm l}) =
\int\!\! dt \, e^{i(\omega_{\rm l}+n\Omega) t}\, \langle e^{iH_F t}  R^{F,\dagger}_{\mathbf q,n} e^{-iH_F t}R^F_{\mathbf q,n} \rangle_F 
\label{whshsswshhha}
\end{align}
is the dynamical structure factor of the scattering operator $R^F_{\mathbf q,n}$. (ii)  F-XRD corresponds to 
additionally 
selecting $\omega_{\rm l}=-n\Omega$. If the system is in a Floquet state, Eq.~\eqref{whshsswshhha} then reduces to $I_{\mathbf q,n}\propto |\langle R^F_{\mathbf q,n}\rangle|^2$, providing a direct probe of the $n$th Floquet component of the scattering operator. (iii) For subcycle-resolved probes, which will not be discussed here, additional oscillatory contributions $\sim e^{i(n-m)\Omega t_p}$ appear as a function of the probe time $t_p$, reflecting interference between different Floquet components $R_{\mathbf q,n}$ and $R_{\mathbf q,m}$. 

What remains is to determine the scattering operator in the presence of a laser field. In general, $R_{\mathbf q}=\sum_{\mathbf r} e^{i\mathbf r\mathbf q} R_{\mathbf r}$ with
\begin{align} 
R_{\mathbf  r}(t)
&=
\mathcal M_\mathbf r\!\!
\int_{0}^{\infty} 
\!\!\!d\tau  \,  
e^{-z\tau }
U(t,t+\tau) c_{\mathbf r} U_{ \mathbf r}(t+\tau,t)  c_{\mathbf r}^\dagger
\label{wgehsssa1main}
\end{align}
for the RIXS operator projected to the valence bands. 
From right to left, this operator describes the creation of an electron in the valence band at site $\mathbf r$ (omitting spin indices for simplicity), the time-evolution $ U_{ \mathbf r}$  of the valence states in the presence of a local potential $-U_c c_{\mathbf r}^\dagger c_{\mathbf r}$
at $\mathbf r$ due to the core hole, the annihilation of the core valence pair, and the back-propagation with the valence time-evolution operator $ U$.
The dipole matrix elements for the X-ray excitation are $\mathcal M_\mathbf r$, and  $e^{-z\tau }$, with $z=\Gamma-i(\omega_{\rm o}-\epsilon_c)$, is an exponential damping with the inverse core-hole lifetime $2\Gamma$  and the detuning $\omega_{\rm o}-\epsilon_c$ of the outgoing X-ray energy from the bare core level energy (i.e., the resonance condition is $z=\Gamma$).

For the numerical evaluation, one can insert Eq.~\eqref{wgehsssa1main} into Eq.~\eqref{eq:rixs_formulasfga}, which gives rise to a numerically costly  integral over a four-point correlation function \cite{Chen2019}. To obtain a simplified expression, we combine Floquet theory with the ultrashort core-hole lifetime approximation (UCA) \cite{vandenBrink2006, Ament2007}. In brief, the evolution operators in Eq.~\eqref{wgehsssa1main} can be expanded for short $\tau$ due to the cutoff provided by $\Gamma$,
and the kick operator $K(t)$ in Eq.~\eqref{gfwehjaa} is obtained within a high-frequency expansion. This leads to a systematic expansion of $R^F_{\mathbf q,n}$ in $1/\Omega$ and $1/\Gamma$ (or $1/z$). The leading contribution to the main band ($n=0$) simply measures the hole density,  
\begin{align}
 R_{\mathbf q,n=0}^F =
\frac{1}{z}\sum_{\mathbf r} e^{i\mathbf q\cdot \mathbf r}
\mathcal M_{\mathbf  r}c_{\mathbf r} c_{\mathbf r}^\dagger.
\label{shsxkskalas}
\end{align}
and is insensitive to the laser drive. To get the  $n\neq 0$ sidebands in laser-dressed scattering, one must go beyond the leading order in $1/\Gamma$.  Here we summarize the central result for a general lattice model (for details of the derivation see Appendix), where the laser vector potential $\mathbf A(t)$ is introduced via a Peierls phase in the hopping. To express the result, it is convenient to  introduce operators
\begin{align}
B_{\ell,\pm}(\mathbf R)=
\frac{s_\pm}{2}
\left(
c^\dagger_{\mathbf R+\mathbf r_\ell}c_{\mathbf R+\mathbf r'_\ell}
\pm
c^\dagger_{\mathbf R+\mathbf r'_\ell}c_{\mathbf R+\mathbf r_\ell}
\right)
\end{align}
for each directed bond $\ell$ connecting sites $\mathbf r'_\ell$ to $\mathbf r_\ell=\mathbf r'_\ell+\boldsymbol\delta_\ell$ relative to the unit cell $\mathbf R$. With $s_+=1$ and $s_-=i$, $B_{\ell,\pm}$ correspond to bond-density ($+$) and bond-current ($-$) operators, respectively.  The nearest neighbor hopping Hamiltonian is simply given by $H_{\rm nn} = -2J\sum_{\mathbf R,\ell}  B_{\ell,+}(\mathbf R)$. For a harmonic drive with $\mathbf A(t)=\mathrm{Re}(\mathbf A_0 e^{-i\Omega t})$, we find that the scattering operator for $n\neq0$  probes a combination of bond-density and bond-current operators, with relative weight determined by the drive amplitude, polarization, and scattering geometry:
\begin{align}
R^F_{\mathbf q,n}
\!=\!
\frac{Js_n}{z(z+in\Omega)}
\sum_{\ell,\pm}
\mathcal J_{n,\ell}
\mathcal R_{n,\pm}\,
F_{\mathbf q,\ell,\pm}
B_{\ell,\pm(-1)^n}(\mathbf q).
\label{eq:R_bond_decomp}
\end{align}
Here $s_n=1 (-i)$ for even (odd) $n$, and the form factors contain the drive-dependent  $\mathcal J_{n,\ell}=\mathcal J_n(|\mathbf A_0\boldsymbol\delta_\ell|)i^ne^{in\arg(\mathbf A_0\boldsymbol\delta_\ell)}$ (with the $n$th Bessel function $\mathcal J_n$), a geometry factor $F_{\mathbf q,\ell,\pm}=\mathcal M_{\mathbf r_\ell}e^{-i\mathbf q\mathbf r_\ell}\pm\mathcal M_{\mathbf r_\ell'}e^{-i\mathbf q\mathbf r'_\ell}$, and an overall factor $\mathcal R_{n,+}=i$ and  $\mathcal R_{n,-} = (-1)^{n+1} \left( 1+\frac{2z}{in\Omega} \right)$, which is distinguishes bond and current contributions.

\begin{figure}[t]
\centerline{\includegraphics[width=0.9\columnwidth]{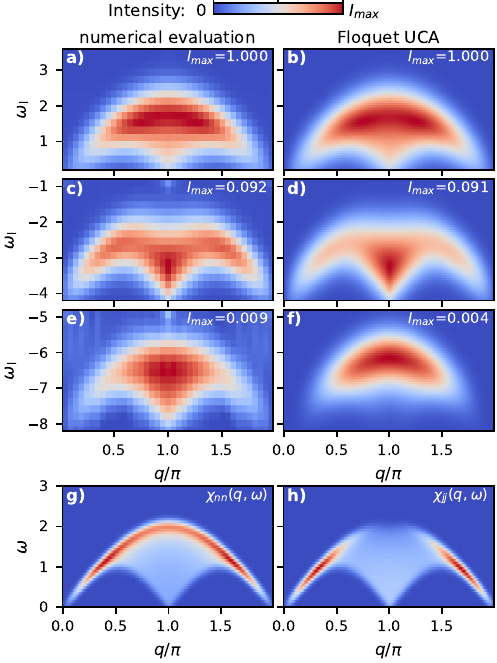}}
\caption{ (a-f)
RIXS signal for the driven 1D chain ($A_0=1.5$, $\Gamma=1$, $\Omega=4$,  $U_c = 2$). 
Results in different spectral regions corresponding to the main Floquet sideband $n=0$ (a,b) and the sidebands  $n=1$ (c,d) and $n=2$ (e,f)  on the energy gain side ($\omega_{\rm l}<0$) are shown with a separate color scale (blue to red for $I\in[0,I_{\rm max}]$, with $I_{\rm max}$ shown by the label). (a,c,e) show the full numerical evaluation of Eqs.~\eqref{eq:rixs_formulasfga} and \eqref{wgehsssa1main}, (b,d,f) the analytical result [Eqs.~\eqref{whshsswshhha} and \eqref{eq:R_bond_decomp}]. In both cases, intensity is integrated over the X-ray frequency $\omega_{\rm o}$; a plot at fixed $\omega_{\rm o}$ would give a similar agreement. (g) density-density correlation, and (h) current-current correlation for 1D chain [with reduced hopping $\mathcal J_0(A_0)$]. }
\label{fig1}
\end{figure}

{\em One-dimensional benchmark ---} For benchmarking, we consider a one-dimensional chain with hopping $J=1$ and  a monochromatic drive $A(t)= A_0\cos(\Omega t)$. With one bond $\ell$ per unit cell,  Eq.~\eqref{eq:R_bond_decomp} becomes 
\begin{align}
R^F_{n,q}  \!=\! 
\frac{i^{n} \mathcal J_{n}(A_0)}{z n\Omega}
\sum_{k} c_{k} ^\dagger c_{k+q}
\left(
m_{k}-\frac{z}{z+in\Omega} m_{k+q}
\right)\!,
\label{materfs}
\end{align}
with $m_k = \epsilon(k)$ and 
$m_k=iv(k)=i\partial_k \epsilon(k)$ for even and odd $n$, respectively. This shows that odd Floquet sidebands can measure the structure factor related to current-current fluctuations, in particular in the limit $\Omega\gg \Gamma$, where the second term in \eqref{materfs} vanishes. [Note that the general result \eqref{eq:R_bond_decomp} spans the limit $\Omega\gg \Gamma$, where the core excited state experiences several laser cycles, and $\Gamma \gg \Omega$, where the core hole decays on a sub-cycle timescale.]

Figure~\ref{fig1} shows the RIXS intensity for a drive with $A_0=1.5$ and $\Omega=4$,  $\Gamma = 1$ and core-valence interaction $U_c=2$.
The probe envelope $s(t)=e^{-t^2\sigma^2}$ with $\sigma=0.45$ spans few cycles of the laser drive. Different rows of the figure show the signal in the spectral range of the main signal (a,b), and the first (c,d) and second (e,f) Floquet sideband on the energy gain side. Right and left panels compare the Floquet UCA with scattering operator  \eqref{materfs}, and a direct numerical evaluation of Eqs.~\eqref{eq:rixs_formulasfga} and \eqref{wgehsssa1main}, respectively, with very good agreement. Overall the range of excitations reflects the particle hole continuum of the one-dimensional chain (shifted by $n\Omega$). However, while the $n=0$ band  resembles the structure factor for density-density correlations $\chi_{nn}(q,\omega)$ (panel g), the $n=1$ band is much closer to the current-current correlations $\chi_{jj}(q,\omega)$ (h); e.g., the weight at $q=\pi$ for the maximal $\omega$, which comes from scattering from states $k=0$ to $k=\pi$, vanishes because $v(k=0)=v(\pi)=0$.  We observe that the numerical benchmark is better for the $n=1$ band compared to the $n=2$ band, which is likely due to contributions $\sim (\Omega,\Gamma)^{-3}$ not included in Eq.~\eqref{eq:R_bond_decomp}. Further benchmarks for the regime  $\Gamma\gg \Omega$ are presented in the Appendix.

\begin{figure}[tbp]
\centerline{\includegraphics[width=0.99\columnwidth]{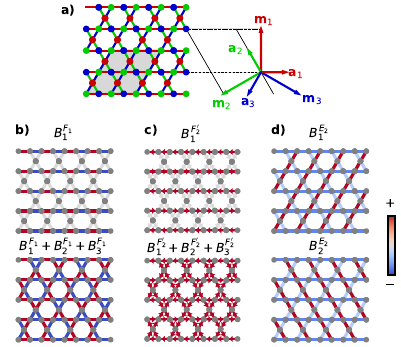}}
\caption{
(a) Kagome lattice, showing the three nearest-neighbor bond directions $\mathbf a_{1,2,3}$, the enlarged unit cell of the symmetry-broken state (shaded), and the ordering wave vectors $\mathbf M_q$. (The indicated vectors $\mathbf m_q$ are parallel to $\mathbf M_q$, with $|\mathbf m_q|=2\pi/|\mathbf M_q|$). (b)–(d) Symmetry-adapted bond patterns $B^{D}_j$ for selected irreducible representations $D=F_1$ (BO), $D=F_2'$ (LCO) and $D=E_2$ (nematic). The color indicates the sign and magnitude of the bond expectation value, while bonds with (without) arrows represent current ($B_{\ell,-}$) and real bond ($B_{\ell,+}$) contributions, respectively.  For the three-dimensional representations $F_1$ and $F_2'$, the upper panels show a single component $j=1$, while the lower panels show the equal superposition of all three components, related by $120^\circ$ rotations. For the nematic order $E_2$ (d), both components $j=1,2$ are shown.}
\label{fig2}
\end{figure}

{\em Probing charge orders on the Kagome lattice ---}  Floquet X-ray scattering is particularly suited to probe instabilities involving  bond and current fluctuations at finite momentum. As a representative example,  we consider charge-ordered states on the Kagome lattice (KL), which serves as a  minimal model for Kagome metals. The KL is a triangular lattice with three sublattices and three nearest-neighbor bond directions $\mathbf a_{1,2,3}$ rotated by $120^\circ$ (Fig.~\ref{fig2}a). At a van Hove filling, Fermi surface hotspots are nested by three wave vectors $\mathbf M_\alpha \perp \mathbf a_\alpha$ (see Fig.~\ref{fig2}a), leading to a  variety of charge density wave instabilities \cite{Fu2024, Wagner2023, Park2021,  Christensen2021, Tan2021,Denner2021}. We focus on orders at these vectors $\mathbf q=\mathbf M_\alpha$, which  imply a doubling of the  unit cell, and work in the enlarged unit cell with $12$ atoms and $24$ nearest-neighbor bonds $B_{\ell}$. Order parameters include conventional density order, characterized by modulations of the site-resolved density $\rho_a=\langle c_{\mathbf R,a}^\dagger c_{\mathbf R,a}\rangle$, bond order (BO), characterized by modulations of $\psi_\ell=\langle B_{\ell,+}\rangle$, and loop-current order (LCO), associated with finite bond currents $j_\ell=\langle B_{\ell,-}\rangle$. According to Eq.~\eqref{eq:R_bond_decomp}, the F-XRD signal $|\langle R^F_{\mathbf q,n}\rangle|^2$  at the Bragg points $\mathbf q=\mathbf M_q$  directly probes a combination of these order parameters. For $n\neq0$, we can write 
\begin{align}
\langle R^F_{\mathbf M_q,n} \rangle \equiv	
\sum_{\ell}
\left(
\mathcal R_{q,n;\ell,+}\, \psi_\ell
+
\mathcal R_{q,n;\ell,-}\, j_\ell
\right),
\label{eq:kagome_signal}
\end{align}
with form factors $\mathcal R_{q,n;\ell,\pm}$   that follow directly from Eq.~\eqref{eq:R_bond_decomp} and depend on the laser  polarization and scattering geometry. In contrast, the $n=0$ component predominantly probes the density order [Eq.~\eqref{shsxkskalas}]. Floquet sidebands therefore provide sensitivity to bond order beyond conventional diffraction.  

To systematically investigate the fingerprint of the ordered states in the F-XRD signal, we decompose the bond operators into  symmetry-adapted combinations $B^{D}_{j} = \sum_{\ell} U^{D}_{j,\ell} B_{\ell,\pm}$ that transform according to irreducible representations $D$  of the Kagome space group \cite{Wagner2023}. The expectation values  $\langle B^{D}_j\rangle$  ($j=1,...,d$) for a $d$-dimensional representation define $d$-dimensional order parameters.  Examples are shown in Fig.~\ref{fig2}: The  characteristic star-of-David BO pattern belongs to the representation $D=F_1$ (Fig.~\ref{fig2}b), the  loop-current order $D=F_2'$ has alternating currents along the  nearest-neighbor chains (Fig.~\ref{fig2}c) (the prime in the label indicates a time-reversal odd order), and the two-dimensional representation  $E_2$ corresponds to a nematic BO that does not break translational  symmetry (Fig.~\ref{fig2}d). A full symmetry classification is given in Ref.~\cite{Wagner2023} and recapitulated in the Appendix.

\begin{figure}[tbp]
\centerline{\includegraphics[width=0.99\columnwidth]{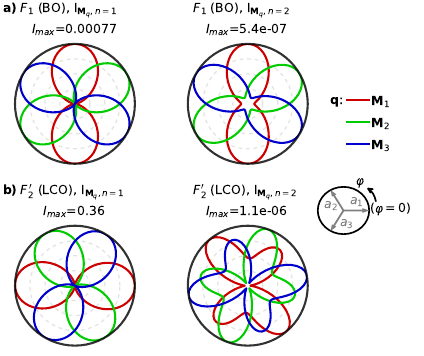}}
\caption{F-XRD intensity $I_{\mathbf q,n}(\varphi)$ as a function of the polarization angle $\varphi$ for selected ordered states on the Kagome lattice. The polar plots show the intensity in the first ($n=1$) and second ($n=2$) Floquet sidebands for the three Bragg vectors $\mathbf q=\mathbf M_{1,2,3}$. The polarization direction $\mathbf A_0=A_0(\cos\varphi,\sin\varphi)$ is defined relative to the crystallographic axes $\mathbf a_1$ (legend on the right). For each panel, the  maximum intensity $I_{\rm max}$ is indicated in the label; $I=1$ corresponds to the intensity of the Bragg peak intensity in the conventional $n=0$ band. The LCO $F_2'$ exhibits an  angular pattern in the first Floquet sideband with intensity maxima for $\mathbf A_0\perp \mathbf q$, while BO $F_1$ yields an orders of magnitude weaker response with a maxima along $\mathbf A_0 \,|\!|\, \mathbf q$. Higher harmonics ($n=2$) are significantly suppressed. Parameters are $A_0=0.5$, $\Gamma=1$,  $U_c = 2$,  $\Omega=4$, $z=\Gamma$ (at resonance).}
\label{fig3}
\end{figure}

The polarization dependence of the form factors in Eq.~\eqref{eq:kagome_signal} is constrained by spatial symmetries and can provide a characteristic fingerprint of the underlying  order parameter.  To illustrate this, we consider the leading instabilities proposed in Ref.~\cite{Fu2024} for spinless fermions with nearest- and next-nearest-neighbor interactions, namely the star-of-David bond order  ($F_1$) and the loop-current order ($F_2'$) shown in Fig.~\ref{fig2}b and c. By adding a mean-field term $H_{D} =  \Delta \sum_{j} B^D_{j}$ to the hopping Hamiltonian, we prepare symmetry-broken states and compute their Floquet scattering signal using Eq.~\eqref{eq:kagome_signal}, for a linearly polarized drive $\mathbf A_0 = A_0 (\cos\varphi \, \hat x + \sin\varphi \, \hat y)$. Figure~\ref{fig3} shows the resulting intensity $I_{q,n}(\varphi) = |\langle R^F_{\mathbf M_q,n} \rangle|^2$ as a function of the polarization angle for the BO $F_1$ and LCO $F_2'$. One finds that first Floquet sideband at the Bragg peak $\mathbf q=\mathbf M_q$ exhibits intensity maxima for $\mathbf A_0 \parallel \mathbf M_q$ and $\mathbf A_0 \perp \mathbf M_q$ for the BO $F_1$ and the LCO $F_2'$, respectively, thus allowing to distinguish between the two states. The three components of the three-dimensional order parameters correspond to the three wave vectors $\mathbf M_q$ and are probed  independently, such that the angular pattern should persist in multi-domain samples.

One can understand this selection rule from spatial symmetry: $\langle R^F_{\mathbf M_q,n=1}\rangle$ vanishes for  $\mathbf A_0 \parallel \mathbf M_q$ ($\mathbf A_0 \perp \mathbf M_q$) if the $\mathbf M_q$ component of the underlying order parameter is odd (even) under a mirror plane parallel to $\mathbf M_q$ (this is demonstrated for all order parameter symmetries in the Appendix). We remark that, in general, a bond order field can also induce a density modulation, and hence lead to a signal at the conventional $n=0$  Bragg peaks at $\mathbf M_q$. The intensity $I_{q,0}=|\langle R^F_{\mathbf M_q,n=0}\rangle |^2$ is taken as a reference intensity for the data in Fig.~\ref{fig3}. However, to leading order, $I_{q,0}$ has no angular dependence and can therefore  not distinguish conventional density order from BO or LCO.  Furthermore, we note that time-reversal symmetry relates scattering amplitudes at $(\mathbf q,n)$ and $(-\mathbf q,-n)$, and, therefore, does not lead to a separate selection rule when analyzing a single Floquet band. Only for $\mathbf q=0$ (relevant, e.g., for probing translationally invariant nematic orders), sidebands for odd $n$ can appear only for nonzero currents $\langle B_{\ell -}\rangle \neq0$, because the form factor $F_{\mathbf q=0,\ell,-}$ in Eq.~\eqref{eq:R_bond_decomp} vanishes. For $\mathbf q\neq0$, odd sidebands are present also for pure real BO. 

\begin{figure}[tbp]
\centerline{\includegraphics[width=0.99\columnwidth]{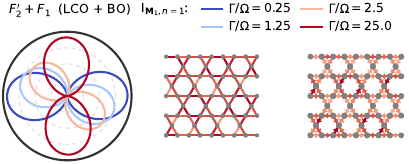}}
\caption{As Fig.~\ref{fig3}, but for a superposition of real BO and LCO, and for different ratios $\Gamma/\Omega$ ($A_0=0.5$, $\Omega=4$, $z=\Gamma$). Only the signal for $\mathbf q=\mathbf M_1$ is shown, the other Bragg peaks are obtained by $120^\circ$ rotation. For large $\Gamma/\Omega$, the intensity becomes dominated by the real bond order, with maxima along $\mathbf A_0 |\!| \mathbf q$. The bond and current pattern is shown in the right insets. [More precisely, the mean-field is $\sum_{j} (B^{F_1 (2)}_j + B^{F_2'}_j/2)$ with the basis functions $B^{D}_j$ given in the Appendix.]  }
\label{fig4}
\end{figure}

On absolute scales, the intensity for $n=1$ is a few orders of magnitude weaker than the elastic signal, but should be experimentally measurable (see \cite{Ornelas-Skarin2026}). The required intensities $A_0\sim0.5$ correspond to a field strength of order of $1$MV/cm, assuming a frequency $\Omega=0.2$eV 
for which sidebands can still be resolved with state-of-the-art resolution, and a typical lattice constant $a=5$\AA {} \cite{DiSante2026}. Yet, as it is sufficient to observe the system over only few driving cycles (see also the benchmark in Fig.~\ref{fig1}), this should be experimentally  feasible. Moreover, the Floquet sidebands can be measured on the energy gain side  ($\omega_{\rm l}<0$),   where they are free of background signals. However, higher Floquet sidebands  $|n|\ge 2$ are probably hard to detect experimentally ($I_n \sim |A_0|^{2n}$ for small $A_0$).  

In addition to the polarization dependence, the relative weight of BO  and LCO in the Floquet signal is controlled by the prefactors $\mathcal R_{n,\pm}$ in Eq.~\eqref{eq:R_bond_decomp}, which depend on the ratio between the drive frequency $\Omega$ and the resonant factor $z=\Gamma -i (\omega_{\rm o}-\epsilon_{\rm c})$. For $|z|\gg\Omega$,  $|\mathcal R_{n,-}| \gg |\mathcal R_{n,+}|$,  and the signal at $n=1$ is  dominated by real bond order. Otherwise, both LCO and BO contribute similarly. The coexistence of multiple order parameters can lead to a characteristic rotation of the polarization pattern as a function of $\Omega/|z|$. This is illustrated in Fig.~\ref{fig4} where we artificially prepare a mixture of a BO and LCO (the corresponding pattern in shown in the inset), and display the angular intensity pattern for different ratios of $\Omega/\Gamma$ (at resonance $z=\Gamma$). As expected, for $\Gamma\gg \Omega$, the signal from the $F_1$ bond order, with maxima for $\mathbf A_0 |\!| \mathbf q$, dominates. Similarly, a rotation of the pattern will appear as a function of the detuning from the resonance $z=\Gamma$.

{\em Conclusion ---} We have developed a  simplified framework to describe resonant Floquet X-ray scattering, by extending the expansion in the core hole lifetime $1/\Gamma$ to periodically driven states. We find that the resonant scattering operator for the Floquet sidebands provides direct sensitivity to bond-resolved degrees of freedom. In this way, it can be used to probe bond and current orders as well as fluctuations which are only indirectly accessible through conventional scattering. Applying this framework  to charge-ordered states on the Kagome lattice, we showed that different symmetry-breaking  patterns exhibit distinct fingerprints in the Floquet signal.  Due to its time-resolved character, the probe could also be used applied to transient nonequilibrium phases in these systems \cite{Grandi2024}.  These results establish Floquet X-ray diffraction as a promising tool to probe  unconventional electronic order.   Beyond diffraction, Floquet  RIXS offers the possibility to access not only static order but also current fluctuations, potentially enabling the detection of fluctuating loop-current correlations in systems such as the cuprates.

{\em Acknowledgements  ---}
We thank Jose Mendez-Guerra for discussions in the initial phase of the project, as well as Kalyani Chordiya. Funding is acknowledged through the Deutsche Forschungsgemeinschaft through OPTIMAL-FOR5750 -  531215165 (Project P1) and through the Cluster of Excellence „CUI: Advanced Imaging of Matter“ of the Deutsche Forschungsgemeinschaft (DFG) – EXC 2056 – project ID 390715994.

%


\onecolumngrid
\begin{appendix}
\allowdisplaybreaks

\counterwithout{equation}{section}
\counterwithout{figure}{section}
\counterwithout{table}{section}

\section{ Xray scattering formalism on periodically driven systems: Derivations}

\setcounter{equation}{0}
\renewcommand{\theequation}{S\arabic{equation}}
\setcounter{figure}{0}
\renewcommand{\thefigure}{S\arabic{figure}}
\setcounter{table}{0}
\renewcommand{\thetable}{S\arabic{table}}

\subsection{Scattering operators}
\label{ssec:scattering_ops}

As in Ref.~\cite{Chen2019}, we consider X-ray scattering in a minimal model with a single localized core orbital per lattice site $\mathbf r$, neglecting inter-site core hopping. The annihilation operators for valence and core electrons are denoted by $c_{\mathbf r}$ and $f_{\mathbf r}$, respectively,
with the corresponding one-particle states being $|w_{\mathbf r}^c\rangle$ and $|w_{\mathbf r}^f\rangle$.	The Hamiltonian reads
\begin{align}
H_{\mathrm{tot}} = H(t) + \epsilon_c \sum_{\mathbf r} f_{\mathbf r}^\dagger f_{\mathbf r} + H_{c\text{-}v},
\label{ppappa}
\end{align}
where $H(t)$ describes the (possibly time-dependent) valence-band Hamiltonian, $\epsilon_c$ is the core-level energy, and
\begin{align}
H_{c\text{-}v} = U_c \sum_{\mathbf r} (f_{\mathbf r}^\dagger f_{\mathbf r}-1)\, c_{\mathbf r}^\dagger c_{\mathbf r}
\end{align}
accounts for the local Coulomb interaction between the core hole and valence electrons
in the presence of a core hole. We assume the core level is initially fully occupied. For notational simplicity, spin indices are suppressed and a single valence band is considered; both can be reinstated without affecting the qualitative results. The expression for time-resolved RIXS between incoming and outgoing photons with energies $\omega_{\rm i}$ and $\omega_{\rm o}$ and momentum transfer $\mathbf q$ is obtained by time-dependent perturbation theory in the dipolar light--matter interaction, yielding \cite{Chen2019,Eckstein2021RIXS}
\begin{align} 
	I_{\mathbf q}(\omega_\text{i}, \omega_\text{o}) 
	= 
	\sum_{\mathbf r,\mathbf r'} e^{i\mathbf q \cdot (\mathbf r-\mathbf r')}
	\int_{t_0}^\infty dt_2 \int_{t_0}^{t_2} dt_1 
	\int_{t_0}^\infty dt_2' \int_{t_0}^{t_2'} dt_1' 
	\, s(t_1) s(t_1') \,\,\,\,\,\,\times
	\nonumber\\
	\times \,\,\,e^{i\omega_{\text{i}}(t_1'-t_1)} 
	e^{-i\omega_{\text{o}}(t_2'-t_2)} 
	e^{-\Gamma (t_2-t_1)}  
	e^{-\Gamma (t_2'-t_1')} 
	\langle P_{\mathbf r'}(t_1') P_{\mathbf r'}^\dagger(t_2') P_{\mathbf r}(t_2) P_{\mathbf r}^\dagger(t_1) \rangle.
	\label{app:rixs_formula_ME}
\end{align}
Here, $P_{\mathbf r}^\dagger = M^{{\rm i}/{\rm o}}_{\mathbf r} c^\dagger_{\mathbf r} f_{\mathbf r}$ are dipolar transition operators, with matrix elements 
$M_\mathbf r^{{\rm i}/{\rm o}} \sim \langle w^c_\mathbf r | \hat{\mathbf r} \cdot \boldsymbol{\epsilon}^{{\rm i}/{\rm o}} | w^f_\mathbf r \rangle$ 
of the dipole operator projected along the polarization $\boldsymbol{\epsilon}^{{\rm i}/{\rm o}}$ of the incoming and outgoing photons. The function $s(t)$ denotes the probe envelope (taken to be Gaussian below), $\Gamma$ 
sets  the inverse core-hole lifetime, and $t_0$ is an initial time before the probe. Assuming translational invariance, the integrand depends only on $\mathbf r - \mathbf r'$. The time-dependent expectation value can be written explicitly as
\begin{align}
\langle  \,
\mathcal U(t_0,t_1'+\tau')
\,P_{\mathbf r'}\,
\mathcal U(t_1',t_1'+\tau')
\,P_{\mathbf r'}^\dagger 
\,\mathcal U(t_1'+\tau',t_1+\tau)
P_{\mathbf r}\,
\mathcal U(t_1+\tau,t_1)
\,P_{\mathbf r}^\dagger  
\,\mathcal U(t_1,t_0) 
\rangle_0,
\end{align}
where  we introduced the parametrization $t_2 = t_1 + \tau$ and $t_2' = t_1' + \tau'$; $\langle \cdots \rangle_0$ denotes the expectation value in the initial state, and $\mathcal U(t',t)$ is the full evolution operator including both the valence dynamics and the core--valence interaction. This expression describes the overlap between a state with a core hole created at site $\mathbf r$ at time $t_1$ and annihilated at $t_1+\tau$, and a corresponding process at site $\mathbf r'$ at times $t_1'$ and $t_1'+\tau'$. Since the core-hole number is conserved during the intermediate evolution, we can restrict the dynamics to the valence sector, incorporating the core-hole potential as a local impurity potential. Denoting by $U_{\mathbf r}(t',t)$ and $U(t',t)$ the valence evolution operators with and without the core-hole potential
at site $\mathbf r$, respectively, the expectation value becomes
\begin{align}
&\langle  
U(t_0,t_1'+\tau')
\,c_{\mathbf r'}\,
U_{\mathbf r'}(t_1',t_1'+\tau')
\,c_{\mathbf r'}^\dagger
U(t_1'+\tau',t_1+\tau)
\,
c_{\mathbf r}\,
U_{\mathbf r}(t_1+\tau,t_1)
\,c_{\mathbf r}^\dagger
\,
U(t_1,t_0)
\rangle_0 
\, e^{i(\tau-\tau')\epsilon_c}.
\end{align}
The last phase factor originates from the bare core-level energy [Eq.~\eqref{ppappa}]. The intensity~\eqref{app:rixs_formula_ME} can then be rewritten as
\begin{align} 
I_{\mathbf q}(\omega_\text{i}, \omega_\text{o}) 
&= 
\sum_{\mathbf r, \mathbf r'} e^{i\mathbf q \cdot (\mathbf r-\mathbf r')}
\int dt_1 dt_1' \, s(t_1) s(t_1')  
e^{i\omega_{\text{l}}(t_1'-t_1)} 
\langle R_{\mathbf r'}^\dagger(t_1') R_{\mathbf r}(t_1) \rangle,
\label{rixs_formulasgsaz}
\end{align}
with the loss energy $\omega_{\rm l}=\omega_{\rm i}-\omega_{\rm o}$, and the RIXS operators
\begin{align} 
R_{\mathbf r}(t)
&=
\mathcal M_{\mathbf r}
\int_{0}^{\infty} d\tau    
\, e^{-z \tau}
U(t,t+\tau)\, c_{\mathbf r}\, U_{\mathbf r}(t+\tau,t)\, c_{\mathbf r}^\dagger,
\label{wgehsssa}
\end{align}
where $\mathcal M_{\mathbf r}=M_{\mathbf r}^{\rm i} M_{\mathbf r}^{\rm o}$ and $z=\Gamma - i(\omega_{\text{o}}-\epsilon_{\rm c})$.  

The factor $z$ therefore incorporates a time cutoff due to the core hole decay, and a phase factor due to the detuning $\omega_{\text{o}}-\epsilon_{\rm c}$ of the X-ray energy from the bare core valence transition. As we are interested in resonant diffraction, we will mostly analyze the results {\em at resonance}, which we define such that $\omega_{\text{o}}-\epsilon_{\rm c}=0$ and thus $z=\Gamma$. Experimentally, one would usually tune the ingoing frequency $\omega_{\rm i}$. For conventional diffraction, $\omega_{\rm i} = \omega_{\rm o}$ , such that the resonant condition $z=\Gamma$ implies $\omega_{\text{i}}=\epsilon_{\rm c}$.  For the scattering into Floquet sidebands with $\omega_{\rm o} = \omega_{\rm i}+n\Omega$ (see below), resonance $z=\Gamma$ would simply require to shift the incoming energy with respect to the conventional resonance $\epsilon_{\rm c}$: $\omega_{\rm i} = \epsilon_{\rm c}-n\Omega$.

\subsection{Scattering formalism for time-periodic systems}

In a periodically driven system, the valence evolution operator can be written in Floquet form as
\begin{align} 
U(t,t') = e^{-iK(t)} e^{-i(t-t')H_F} e^{iK(t')},
\label{sgsflosgs}
\end{align}
where $K(t)$ is a time-periodic kick operator and $H_F$ the time-independent Floquet Hamiltonian \cite{Bukov2015}. Inserting this into Eq.~\eqref{rixs_formulasgsaz}, we obtain
\begin{align} 
	I_{\mathbf q}(\omega_\text{i}, \omega_\text{o}) 
	&= 
	\sum_{\mathbf r, \mathbf r'} e^{i\mathbf q \cdot (\mathbf r-\mathbf r')}
	\int dt_1 dt_1'  \, s(t_1) s(t_1') e^{i\omega_{\text{l}}(t_1'-t_1)}
	\langle 
	e^{-i(t_0-t_1')H_F} 
	R_{\mathbf r'}^{F}(t_1')^\dagger
	e^{-i(t_1'-t_1)H_F}
	R^F_{\mathbf r}(t_1)
	e^{-i(t_1-t_0)H_F}
	\rangle_F,
	\label{app:rixs_formula_ME08ucl11}
\end{align}
with the dressed RIXS operators
\begin{align} 
R^F_{\mathbf r}(t) =  e^{iK(t)} R_{\mathbf r}(t) e^{-iK(t)},
\label{gesdfjagfds01}
\end{align}
and the Floquet expectation value $\langle\cdots\rangle_F =  \mathrm{tr}(\cdots \rho_F)$, where $\rho_F=e^{iK(t_0)} \rho_0 e^{-iK(t_0)}$. The operator $R^F_{\mathbf r}(t)$ is periodic in time due to both the kick operator and the intrinsic time dependence of $R_{\mathbf r}(t)$ [see Eq.~\eqref{wgehsssa}]. 
Therefore, it can be expanded in a Fourier series,
\begin{align}
R^F_{\mathbf r}(t) =
\sum_{n}  R^F_{\mathbf r,n}\, e^{-in\Omega t}.
\label{gesdfjagfds}
\end{align}
With this expansion, the RIXS intensity \eqref{app:rixs_formula_ME08ucl11} becomes
\begin{align} 
	I_{\mathbf q}(\omega_\text{i}, \omega_\text{o}) 
	&= 
	\int dt\, dt' \, s(t) s(t')  e^{i\omega_{\text{l}}(t'-t)}
	\sum_{n,n'}
	 e^{i\Omega n' t'} e^{-i\Omega n t}
	\,\,Y_{\mathbf q;n',n}(t',t),
\end{align}
with the dynamical correlation function
\begin{align}
	Y_{\mathbf q;n',n}(t',t)=&
	\langle e^{-i(t_0-t')H_F} 
	\, ( R^F_{n',\mathbf q}  )^\dagger\,
	e^{-i(t'-t)H_F}
	\, R^F_{n,\mathbf q} \,
	e^{-i(t-t_0)H_F}
	\rangle_F,
	\label{eq:rixs_formula_ME08ucl2}
\end{align}
where
\begin{align}
R^F_{n,\mathbf q} = \sum_{\mathbf r} e^{i\mathbf q \cdot \mathbf r} R^F_{\mathbf r,n}. 
\label{eq:R_nq}
\end{align}

We will further assume that the correlation function is stationary, i.e., $Y_{\mathbf q;n',n}(t',t)=Y_{\mathbf q;n',n}(t'-t)$. 
This assumption is justified in two relevant limits: First, it holds if the system is prepared in a Floquet steady state, either via an adiabatic ramp of the drive or through relaxation to a periodic steady state. In these two cases, the density matrix $\rho_F$ is diagonal in the eigenbasis of $H_F$.  Second, stationarity also applies in the perturbative regime of weak driving, where the Fourier components of the scattering operator scale as $R^F_{\mathbf q,n} \sim A_0^n$ with the drive amplitude $A_0$ (see below). To leading order in $A_0$, contributions from a given Floquet sector can then be evaluated using the equilibrium density matrix, effectively replacing $\rho_F$ by the undriven one.  The latter regime is particularly relevant for using Floquet RIXS as a probe of unconventional ordered states, where the drive acts as a weak probe rather than inducing a fully developed Floquet steady state.

Using a Gaussian probe envelope with probe duration $1/\sigma$,
\begin{align} 
s(t) = e^{-\sigma^2(t-t_p)^2} \equiv g_\sigma(t-t_p),
\end{align}
the product $s(t)s(t')$ factorizes as $s(t)s(t') = g_{\sigma/\sqrt{2}}(t_{\rm rel})\, g_{\sqrt{2}\sigma}(t_{\rm av}-t_p)$, in a parametrization using relative and average times $t_{\rm rel} = t' - t$, $t_{\rm av} = \frac{t+t'}{2}$.
Hence,
\begin{align} 
	I_{\mathbf q} (\omega_\text{i}, \omega_\text{o}) 
&= 
\sum_{n,n'} \int dt_{\rm av}\, g_{\sqrt{2}\sigma}(t_{\rm av}-t_p)\, e^{-i\Omega (n-n') t_{\rm av}}
\int dt_{\rm rel}\, e^{i\left(\omega_{\text{l}}+\frac{n+n'}{2}\Omega\right)t_{\rm rel}} \,
g_{\sigma/\sqrt{2}}(t_{\rm rel})\,
Y_{\mathbf q;n',n}(t_{\rm rel}).
\end{align}
Performing the $t_{\rm av}$ integral yields
\begin{align} 
I_{\mathbf q}
(\omega_\text{i}, \omega_\text{o}) 
&\sim 
\sum_{n,n'}
e^{
-\frac{\Omega^2 (n-n')^2}{8\sigma^2}
}
e^{-i\Omega (n-n') t_p}
\int dt_{\rm rel}\, e^{i\left(\omega_{\text{l}}+\frac{n+n'}{2}\Omega\right)t_{\rm rel}} \,
g_{\sigma/\sqrt{2}}(t_{\rm rel})\,
Y_{\mathbf q;n',n}(t_{\rm rel}).
\label{wbsjaisaaasds11}
\end{align}
Equation~\eqref{wbsjaisaaasds11} provides a natural interpolation between cycle-averaged and subcycle-resolved scattering regimes, controlled by the probe duration $1/\sigma$ relative to the drive period $2\pi/\Omega$:
\begin{itemize}
\item
For long probe pulses, $\Omega \gg \sigma $, the Gaussian factor  $\exp\left({-\frac{\Omega^2 (n-n')^2}{8\sigma^2}}\right)$
strongly suppresses contributions with $n \neq n'$, effectively enforcing $n=n'$. In this limit, the signal reduces to a sum over diagonal Floquet sectors,
\begin{align}
I_{\mathbf q}(\omega_\text{i}, \omega_\text{o}) 
\sim 
\sum_n \int dt_{\rm rel}\, e^{i(\omega_{\text{l}}+n\Omega)t_{\rm rel}} 
Y_{\mathbf q;n,n}(t_{\rm rel}),
\label{wbsjaisaaasds12}
\end{align}
which corresponds to a superposition of Floquet sidebands probing the dynamical structure factor associated with $R^F_{\mathbf q,n}$. 
\item
In contrast, for short probe pulses with duration comparable to or shorter than the driving period, $\Omega \lesssim \sigma $, the Gaussian broadening allows for contributions with $n \neq n'$. In this regime, the phase factor $e^{-i\Omega (n-n') t_p}$ in Eq.~\eqref{wbsjaisaaasds11} leads to oscillatory dependence of the signal on the probe time $t_p$, reflecting interference between different Floquet components $R^F_{\mathbf q,n}$ and $R^F_{\mathbf q,n'}$. These off-diagonal terms encode correlations between distinct Floquet sectors and provide, in principle, access to coherences in Floquet space. In this manuscript, we focus on the experimentally easier accessible cycle-averaged regime \eqref{wbsjaisaaasds12}. 
\item
Finally, selecting energy transfers $\omega_{\rm l} \approx -n\Omega$ isolates a Floquet sideband of the resonant X-ray diffraction. In particular, in a pure Floquet state the signal reduces to
\begin{align}
I_{\mathbf q,n} \propto |\langle R^F_{\mathbf q,n} \rangle|^2,
\end{align}
providing a direct probe of the $n$th harmonic of the scattering operator. This relation will be used below to analyze the symmetry of the scattering signal.
\end{itemize}

\subsection{Floquet ultrashort core-hole lifetime expansion (F-UCL)}

To simplify the Floquet scattering operator $R^F_{\mathbf q,n}$ [Eq.~\eqref{gesdfjagfds}], we use that both the inverse core-hole lifetime and the driving frequency $\Omega$ can be large or comparable to the relevant energy scales of the valence-band model; $1/\Gamma$ can be on the order of a few fs in real materials, while $\Omega$ can be chosen sufficiently large to separate the Floquet sidebands. Our goal is therefore to derive a controlled expansion in powers of $1/\Gamma$ (more precisely $1/|z|$) and $1/\Omega$. Physically, this expansion corresponds to a short-time expansion of the intermediate-state evolution in the presence of the core-hole potential and the periodic drive. We derive the expansion to leading nontrivial order in both parameters. In the main text we benchmark this approximation and find remarkably good agreement already for moderately large values $\Gamma=1$ and $\Omega=4$ in a tight-binding model with hopping $J=1$.  While both $1/\Omega$ and $1/\Gamma$ are treated as small parameters, the resulting expressions still interpolate between the limits $\Omega \gg \Gamma$, where the intermediate state evolves over several driving cycles, and the opposite limit $\Omega \ll  \Gamma$, where the effect of the drive on the intermediate state is less pronounced.

The RIXS operator can be simplified in the limit of a short core-hole lifetime (large $\Gamma$), in which case the evolution operators can be expanded as
\begin{align}
U_{(\mathbf r)}(t+\tau,t) 
&=  1
-i\int_{t}^{t+\tau} ds \,H_{(\mathbf r)}(s) 
+\frac{(-i)^2}{2} \int_{t}^{t+\tau} ds_1 ds_2 \,
T_t\big[H_{(\mathbf r)}(s_1) H_{(\mathbf r)}(s_2)\big]
+\cdots.
\end{align}
Upon inserting this into Eq.~\eqref{wgehsssa}, the $n$th order term in this expansion ultimately contributes to order $n+1$ for the RIXS operator due to the integration $\int_{0}^{\infty} d\tau \, e^{-z\tau }(\cdots)$. We obtain the series
\begin{align}
 R_{\mathbf r}(t)
&=
 R^{(0)}_{\mathbf r}(t)
+ 
R^{(1)}_{\mathbf r}(t) + \cdots,
\end{align}
where we keep only the leading orders $n=0,1$:
\begin{align} 
 R_{\mathbf r}^{(0)}(t)
&=
\int_{0}^{\infty} d\tau\, e^{-z\tau }
 c_{\mathbf r}  c_{\mathbf r}^\dagger
=
 \frac{c_{\mathbf r}  c_{\mathbf r}^\dagger}{z},
\label{wgehsssa0-1}
\\
R^{(1)}_{\mathbf r}(t)
&=
i
\int_{0}^{\infty} d\tau\, e^{-z\tau }
\int_{t}^{t+\tau} ds 
\left(
 H(s)\, c_{\mathbf r} 
 -
c_{\mathbf r}  H_{\mathbf r}(s)
\right)
c_{\mathbf r}^\dagger.
\label{wgehsssa1}
\end{align}
For the first-order term we write
\begin{align}
H_{\mathbf r}(s)= H(s) + V_{\mathbf r},
\end{align}
where $V_{\mathbf r}=-U_c c_{\mathbf r}^\dagger c_{\mathbf r}$ is the core-hole potential, assumed time-independent, and $H(s)$ is the valence Hamiltonian. Equation~\eqref{wgehsssa1} then becomes
\begin{align}
R^{(1)}_{\mathbf r}(t)
&=
 R^{(1a)}_{\mathbf r}(t)
+
 R^{(1b)}_{\mathbf r}(t),
\\
 R^{(1a)}_{\mathbf r}(t)
&=
i
\int_{0}^{\infty} d\tau\, e^{-z\tau }
\int_{t}^{t+\tau} ds 
[H(s), c_{\mathbf r} ]c_{\mathbf r}^\dagger,
\label{wgehsssa1a}
\\
R^{(1b)}_{\mathbf r}(t)
&=
-i
\int_{0}^{\infty} d\tau\, e^{-z\tau }
\int_{t}^{t+\tau} ds 
c_{\mathbf r}  V_{\mathbf r}
c_{\mathbf r}^\dagger
=
\frac{c_{\mathbf r}  V_{\mathbf r}
c_{\mathbf r}^\dagger}{iz^2}.
\label{wgehsssa1b}
\end{align}
The operator $R^{(1a)}$ is the leading term that captures the effect of the time-dependent fields in the valence band on the scattering process. At this order, the scattering signal is therefore modified by the driving fields both through the dressing of the RIXS operator and through the modified intermediate-state dynamics.

While Eqs.~\eqref{wgehsssa1a} and \eqref{wgehsssa1b} are valid for a general time-dependent Hamiltonian, further simplifications occur for a periodically driven system. In this case, the Hamiltonian can be written as a Fourier series,
\begin{align}
H(s) = \sum_{n} e^{-i\Omega n s} H_n.
\end{align}
Inserting this into Eq.~\eqref{wgehsssa1a} yields
\begin{align}
R^{(1a)}_{\mathbf r}(t)
&=
\sum_n
i
[H_n, c_{\mathbf r} ]c_{\mathbf r}^\dagger 
\int_{0}^{\infty} d\tau\, e^{-z\tau }
\int_{t}^{t+\tau} ds \,
e^{-isn\Omega}
\nonumber\\
&=
\sum_n
e^{-itn\Omega}
\frac{
i
[H_n, c_{\mathbf r} ]c_{\mathbf r}^\dagger 
}{z(z+in\Omega)}.
\label{wgewaaaz}
\end{align}
To obtain the Floquet scattering operator, the RIXS operator must then be dressed with the kick operator as in Eq.~\eqref{gesdfjagfds01}. We now use the high-frequency expansion. To first order in $\Omega^{-1}$, the Floquet Hamiltonian is \cite{Bukov2015}
\begin{align}
H_F
&=
H_0
+
\sum_{m\neq 0}
\frac{[H_{-m},H_m]}{m\Omega}
+
\mathcal O(\Omega^{-2}),
\label{HFE}
\end{align}
while the kick operator is
\begin{align}
K(t)
&=
-\sum_{m\neq 0}
\frac{H_m}{im\Omega}
\,e^{-im\Omega t}
+
\mathcal O(\Omega^{-2}).
\end{align}
The Floquet RIXS operator \eqref{gesdfjagfds01} then becomes
\begin{align} 
e^{iK(t)}  R_{\mathbf r}(t) e^{-iK(t)} 
&=
R_{\mathbf r}(t)
-i
\sum_{m\neq 0}
\frac{e^{-im\Omega t}}{im\Omega}
[H_m,  R_{\mathbf r}(t)].
\label{gesdfjagfds011}
\end{align}
In this expression, we now  keep only the leading contributions in the expansion in $1/\Gamma$ and $1/\Omega$. For a Floquet sideband $n\neq 0$, there are two contributions:
One from the $n$th Fourier component in Eq.~\eqref{wgewaaaz} inserted into the first term in Eq.~\eqref{gesdfjagfds011}, and one from the commutator $[H_n, R_{\mathbf r}(t)]$ in Eq.~\eqref{gesdfjagfds011}, where the appearing RIXS operator is approximated by the leading-order operator $R^{(0)}_{\mathbf r}$ [Eq.~\eqref{wgehsssa0-1}]. Altogether, this yields
\begin{align} 
R^F_{\mathbf r,n}
&=
\frac{i [H_n, c_{\mathbf r} ]c_{\mathbf r}^\dagger }{z(z+in\Omega)}
-
\frac{i}{z}
\frac{[H_n,  c_{\mathbf r}c_{\mathbf r}^\dagger]}{in\Omega}
\nonumber\\
&=
\frac{i}{z}
\left(
\frac{ [H_n, c_{\mathbf r} ]c_{\mathbf r}^\dagger }{z+in\Omega}
-
\frac{[H_n,  c_{\mathbf r}]c_{\mathbf r}^\dagger +  c_{\mathbf r}[H_n, c_{\mathbf r}^\dagger]}{in\Omega}
\right)
\nonumber\\
&=
\frac{1}{iz}
\frac{1}{in\Omega}
\left(
[H_n, c_{\mathbf r} ]c_{\mathbf r}^\dagger
\frac{ z }{z+in\Omega}
+
c_{\mathbf r}[H_n, c_{\mathbf r}^\dagger]
\right).
\label{ehjheasaaaa}
\end{align}
Reinserting the dipole matrix elements and transforming to reciprocal space,  the Floquet RIXS operator becomes
\begin{align} 
R^F_{\mathbf q,n} 
&=  
\frac{1}{iz}
\frac{1}{in\Omega}
\sum_{\mathbf r} e^{-i\mathbf q\cdot\mathbf r } 
\mathcal M_{\mathbf r}
\left(
[H_n, c_{ \mathbf r} ]c_{ \mathbf r}^\dagger
\frac{ z }{z+in\Omega}
+
c_{  \mathbf r}[H_n, c_{ \mathbf r}^\dagger]
\right).
\label{sllshgehsss33}
\end{align}

This is the central result for the Floquet RIXS operator at nonzero sidebands. One can see that the leading contribution to the sidebands with $n\neq0$ requires going beyond the leading $1/z$ term in the ultrashort core-hole lifetime expansion, which reflects the fact that the electron must spend a finite time in the valence band in order to experience the periodic drive. As a result, the sidebands probe the operators through which the drive couples to the system, such as bond or current operators in a tight-binding model, as further worked out in Section \ref{aecslsshhssas}. For the main sideband $n=0$, the leading contribution is instead simply proportional to the density [Eq.~\eqref{wgehsssa0-1}]. 

\subsection{Expansion of the Floquet scattering operator in bond operators}
\label{aecslsshhssas}

In a driven system, electromagnetic fields couple generically to the kinetic (hopping) terms via Peierls phases, i.e., to bond operators. It is therefore natural to express both the Hamiltonian and the Floquet RIXS operator in a bond-operator basis. We let $\mathbf R$ denote the unit-cell position and use an index $\ell$ to label directed bonds with endpoint inside the unit cell. For each bond, we denote by $\mathbf r_\ell$ and $\mathbf r_\ell' \equiv \mathbf r_\ell-\boldsymbol\delta_\ell$ the end and start points,
and define
\begin{align}
B_{\ell,\pm}(\mathbf R) 
&=
\frac{s_\pm}{2}\left(
c_{\mathbf R+\mathbf r_\ell}^\dagger c_{\mathbf R+\mathbf r_\ell'}
\pm
c_{\mathbf R+\mathbf r_\ell'}^\dagger c_{\mathbf R+\mathbf r_\ell} 
\right),
\end{align}
where $s_+=1$ and $s_-=i$. Then, a translationally invariant hopping Hamiltonian is of the form
\begin{align}
H_{\rm hop} = -2\sum_{\ell} J_\ell\, B_{\ell,+},
\end{align}
where here and in the following 
\begin{align}
B_{\ell,\pm}=
\sum_{\mathbf R}
B_{\ell,\pm}(\mathbf R) 
\end{align}
is the translationally invariant superposition of bond operators in all cells. In the following, for simplicity of notation, we assume all hopping amplitudes $J_\ell$ to be equal, as for symmetry-equivalent nearest-neighbor bonds.

In the presence of a vector potential $\mathbf A(t)$, the hopping terms acquire Peierls phases,
\begin{align}
c_{\mathbf r}^\dagger c_{\mathbf r'}
\;\to\;
c_{\mathbf r}^\dagger c_{\mathbf r'}\,
e^{i(\mathbf r-\mathbf r')\cdot \mathbf A(t)}.
\end{align}
One can therefore introduce the Peierls-dressed bond operators
\begin{align}
\bar B_{\ell}(\mathbf R,t) 
&= 
\frac{1}{2}\left(
c_{\mathbf R+\mathbf r_\ell}^\dagger c_{\mathbf R+\mathbf r_\ell'}
e^{i\mathbf A(t)\cdot\boldsymbol\delta_{\ell}}
+
h.c.\right)
\nonumber\\
&= 
\cos\!\left(\mathbf A(t)\cdot\boldsymbol\delta_{\ell}\right) B_{\ell,+}(\mathbf R) 
+\sin\!\left(\mathbf A(t)\cdot\boldsymbol\delta_{\ell}\right) B_{\ell,-}(\mathbf R).
\end{align}
For a monochromatic drive $\mathbf A(t)= \mathrm{Re}\!\left(e^{-i\Omega t}\mathbf A_0\right)$, we parametrize
\begin{align}
\mathbf A(t)\cdot \boldsymbol\delta_{\ell}
=
\frac{1}{2}e^{-i\Omega t} z_{\ell} + h.c.,
\qquad
z_{\ell}= \mathbf A_0\cdot  \boldsymbol\delta_{\ell}
\equiv |z_{\ell}| e^{i\xi_{\ell}}.
\end{align}
Using the identity 
\begin{align}
\frac{\Omega}{2\pi}\int_0^{2\pi/\Omega} dt \, e^{in\Omega t}
e^{i|z|\cos(\Omega t - \xi_{\ell})}
=
e^{in\xi_{\ell}} i^n \mathcal J_n(|z|),
\label{sgsgwwwza33}
\end{align}
with $\mathcal J_n(x)$ the Bessel function, we obtain
\begin{align}
\cos(\mathbf A(t)\cdot\boldsymbol\delta_{\ell})
&=
\sum_{n:{\rm even}} e^{-in\Omega t}\,
\mathcal J_{n}(|z_{\ell}|)
(ie^{i \xi_{\ell}})^n,
\\
\sin(\mathbf A(t)\cdot\boldsymbol\delta_{\ell})
&=
\sum_{n:{\rm odd}} e^{-in\Omega t}\,
(-i)\mathcal J_{n}(|z_{\ell}|)
(ie^{i \xi_{\ell}})^n.
\end{align}
Introducing the shorthand
\begin{align}
\mathcal J_{n,\ell}= \mathcal J_{n}(|\mathbf A_0\cdot\boldsymbol\delta_{\ell}|)
(ie^{i \xi_{\ell}})^n ,
\label{genjsge}
\end{align}
the dressed bond operators have the Fourier decomposition
\begin{align}
\bar B_{\ell}(\mathbf R,t)
&= 
\sum_{n} e^{-in\Omega t}\, s_n
\mathcal J_{n,\ell}
B_{\ell,(-1)^n}(\mathbf R),
\end{align}
where $s_n=1$ for even $n$ and $s_n=-i$ for odd $n$.
With that, the driven hopping Hamiltonian is
\begin{align}
H(t)=-2J\sum_{\mathbf R,\ell} \bar B_\ell(\mathbf R,t)
=
\sum_{n} e^{-in\Omega t} H_n,
\end{align}
with Fourier components
\begin{align}
H_n
&=
-2J \sum_{\ell,\mathbf R}
s_n\mathcal J_{n,\ell}\,
B_{\ell,(-1)^n}(\mathbf R).
\label{wgzidegejzs171}
\end{align}

We can now evaluate the Floquet RIXS operator starting from Eq.~\eqref{sllshgehsss33}. Given Eq.~\eqref{wgzidegejzs171}, this requires the commutators $[B_{\ell,\pm}(\mathbf R), c_{\mathbf r}]c_{\mathbf r}^\dagger$ and $c_{\mathbf r}[B_{\ell,\pm}(\mathbf R), c_{\mathbf r}^\dagger]$.
Performing the sum over lattice sites and using fermionic commutation relations, the commutators can be expressed in terms of the bond operators themselves,
\begin{align}
\sum_{\mathbf r} e^{-i\mathbf q\cdot\mathbf r}  \mathcal M_{\mathbf r}
[B_{\ell,\pm}(\mathbf R), c_{\mathbf r}]c_{\mathbf r}^\dagger
&=
\frac{1}{2}
e^{-i\mathbf q\cdot\mathbf R} 
\left(
F_{\mathbf q,\ell,+}
B_{\ell,\pm}(\mathbf R)
\mp i\,
F_{\mathbf q,\ell,-}
B_{\ell,\mp}(\mathbf R)
\right),
\\
\sum_{\mathbf r} e^{-i\mathbf q\cdot\mathbf r} \mathcal M_{\mathbf r}\,
c_{\mathbf r}[B_{\ell,\pm}(\mathbf R), c_{\mathbf r}^\dagger]
&=
-\frac{1}{2}
e^{-i\mathbf q\cdot\mathbf R} 
\left(
F_{\mathbf q,\ell,+}
B_{\ell,\pm}(\mathbf R)
\pm i\,
F_{\mathbf q,\ell,-}
B_{\ell,\mp}(\mathbf R)
\right).
\end{align}
where
\begin{align}
F_{\mathbf q,\ell,\pm}
=
\mathcal M_{\mathbf r_\ell} e^{-i\mathbf q\cdot\mathbf r_\ell}
\pm
\mathcal M_{\mathbf r_\ell'} e^{-i\mathbf q\cdot\mathbf r_\ell'}
\end{align}
are geometry-dependent form factors. Inserting these expressions into Eq.~\eqref{sllshgehsss33} and collecting terms, we find
\begin{align}
R_{\mathbf q,n}^F
&=
 \frac{-J}{iz\,in\Omega}
\sum_{\ell}
s_n\mathcal J_{n,\ell}
\Bigg[
F_{\mathbf q,\ell,+}
B_{\ell,(-1)^n}(\mathbf q)
\left(\frac{z}{z+in\Omega} -1 \right)
-i(-1)^n
\left(\frac{z}{z+in\Omega} +1 \right)
F_{\mathbf q,\ell,-}
B_{\ell,(-1)^{n+1}}(\mathbf q)
\Bigg]
\\
&=
-J
\sum_{\ell}
\frac{s_n\mathcal J_{n,\ell} 
}{z(z+in\Omega)}
\Bigg[
iF_{\mathbf q,\ell,+}
B_{\ell,(-1)^n}(\mathbf q)
-(-1)^n
\frac{2z+in\Omega}{in\Omega}
F_{\mathbf q,\ell,-}
B_{\ell,(-1)^{n+1}}(\mathbf q)
\Bigg],
\end{align}
where $B_{\ell,\pm}(\mathbf q)=\sum_{\mathbf R} e^{-i\mathbf q\cdot\mathbf R} B_{\ell,\pm}(\mathbf R)$.
Finally, $R_{\mathbf q,n}^F$ can be written in the compact form used in the main text
[Eq.~\eqref{eq:R_bond_decomp}], 
\begin{align}
R^F_{\mathbf q,n}
=
\frac{-Js_n}{z(z+in\Omega)}
\left(
\sum_{\ell}
\mathcal J_{n,\ell}\,
\mathcal R_{n,+}\,
F_{\mathbf q,\ell,+}\,
B_{\ell,(-1)^n}(\mathbf q)
+
\sum_{\ell}
\mathcal J_{n,\ell}\,
\mathcal R_{n,-}\,
F_{\mathbf q,\ell,-}\,
B_{\ell,-(-1)^n}(\mathbf q)
\right),
\label{hskaxz}
\end{align}
with
\begin{align}
\mathcal R_{n,+}
&=
i ,
\label{shgxshsqq1}
\\
\mathcal R_{n,-}
&=
(-1)^{n+1}
\left(
1+\frac{2z}{in\Omega}
\right),
\label{shgxshsqq2}
\end{align}
where $s_n=1$ for even $n$ and $s_n=-i$ for odd $n$. Thus, each Floquet sideband couples to a specific linear combination of real bond operators 
$B_{\ell,+}$ and bond current operators $B_{\ell,-}$, with coefficients controlled by the drive geometry, the dipole matrix elements, and the ratio of the core-hole decay $\Gamma$ and the driving frequency $\Omega$.

For $\mathbf q=0$, and site-independent dipole matrix elements $\mathcal M_{\mathbf r}$, the form factor reduces to
\begin{align}
F_{\mathbf 0,\ell,s} = 2\mathcal M\,\delta_{s,+},
\end{align}
such that
\begin{align}
R^F_{\mathbf 0,n}
=
2\sum_{\ell}
\mathcal J_{n,\ell}\,
\mathcal R_{n,+}\,
B_{\ell,(-1)^n}.
\label{hskaxz-1}
\end{align}
Hence, at $\mathbf q=0$, the Floquet scattering operator strictly separates contributions from bond-density and bond-current operators into even and odd Floquet sidebands, respectively. For probing unconventional current order at $\mathbf q\neq 0$, however,  a signal in the $n=1$ sideband alone is not sufficient evidence for current order; instead, different orders are distinguished by the angular dependence and polarization dependence of the scattering signal 
for that sideband 
(see main text). 

\subsection{Time-reversal symmetry}

We briefly discuss the transformation of the Floquet scattering operator under time-reversal symmetry. 
For spinless fermions, the time-reversal operator $\mathcal T$ acts as complex conjugation
in combination with 
\begin{align}
\mathcal T\, c_{\mathbf r}\, \mathcal T^{-1} = c_{\mathbf r}.
\end{align}
With this, the bond operators $B_{\ell,+}(\mathbf R)$ ($B_{\ell,-}(\mathbf R)$) are even (odd) under time reversal. Furthermore, under complex conjugation the signs of 
$s_n$ is reversed for odd $n$ (since $s_n=-i$ for odd $n$), and 
\begin{align}
 \mathcal J_{n,\ell} ^*
 =
 \mathcal J_{n}(|\mathbf A_0\cdot\boldsymbol\delta_{\ell}|)
(-ie^{-i \xi_{\ell}})^{n} 
 = (-1)^n \mathcal J_{-n,\ell},
\end{align}
using Eq.~\eqref{genjsge} and $\mathcal J_n(x)=(-1)^n \mathcal J_{-n}(x)$. Equation \eqref{wgzidegejzs171} therefore implies
\begin{align}
\mathcal T H_n \mathcal T^{-1}=(-1)^n H_{-n}.
\end{align}
Both from the general form \eqref{sllshgehsss33} of the Floquet scattering operator and from the explicit expansion \eqref{hskaxz}, we can conclude
\begin{align}
\mathcal T R^F_{\mathbf q,n}(z) \mathcal T^{-1}
=
(-1)^{n+1}R^F_{-\mathbf q,-n}(z^*).
\end{align}
Thus, time-reversal symmetry relates positive and negative Floquet sidebands. As discussed in the main text,  this implies that a scattering experiment that selects a single Floquet sideband $n$ does not separate time-reversal even and odd components. In principle, time-reversal even and odd contributions can be accessed in cycle-resolved measurements, which are sensitive to interference terms between different Floquet components $R^F_{\mathbf q,n}$ and $R^F_{\mathbf q,n'}$. However, for resonant scattering interpreting such interference terms is 
more complicated because different Floquet channels involve different complex factors $z=\Gamma-i(\omega_{\rm o}-\epsilon_{\rm c})$ if the  incoming photon energy is the same for both bands.  Here we therefore focus on individual Floquet sidebands, and leave the investigation of subcycle-resolved resonant scattering to future work.

\subsection{Floquet Kramers-Heisenberg expression}

Although we have derived the Floquet scattering explicitly via a high-frequency and short core-hole lifetime expansion, it is interesting to note that a more general formal expression for driven RIXS can also be obtained.
Here we derive a Floquet extension of the well-known Kramers-Heisenberg formula for equilibrium RIXS, which holds for any core-hole lifetime.

Starting from Eq.~\eqref{app:rixs_formula_ME08ucl11}, one 
can insert a resolution of unity in terms of the eigenstates $\{|m^F\rangle\}$  with energies  $\epsilon^F_m$ of the Floquet Hamiltonian $H_F$ into the expectation value, such that the scattering amplitude in the Floquet description in Eq.~\eqref{app:rixs_formula_ME08ucl11} can be written as
\begin{align}
	I_{\mathbf q}(\omega_i,\omega_o) &= \sum_{m,j} w_j^F \, \left|M_{m,j}(\mathbf q,\omega_i,\omega_o) \right|^2,\\
	M_{m,j}(\mathbf q,\omega_i,\omega_o) &= \sum_n \tilde s(\omega_l-\epsilon_m^F+\epsilon_j^F+n\Omega)\, \langle m^F|R_{n,\mathbf q}|j^F\rangle.
\end{align}
Here, the Fourier series of $R^F_{\mathbf r}(t)$ [Eq.~\eqref{gesdfjagfds}] and its momentum representation [Eq.~\eqref{eq:R_nq}] were inserted. The Fourier transform of the probe envelope is denoted by $\tilde s(\omega)= \int \mathrm dt\, s(t) \,e^{i\omega t}$, and the dressed initial density matrix is 
assumed to be diagonal in  terms of eigenstates of $H_F$, $\rho_F = e^{-iK(t_0)}\rho_0 e^{iK(t_0)}= \sum_j w_j^F | j^F\rangle\langle j^F|$,  which holds automatically if $\rho_0$ is prepared as ensemble of Floquet modes.

Following the notation in Section \ref{ssec:scattering_ops}, where the core Hilbert space has been factored out, one can relate the Floquet Hamiltonian $H_{F,\mathbf r}$ in the intermediate, X-ray excited state (with core-hole at site $\mathbf r$) to the Floquet Hamiltonian $H_F$ before and after the scattering process, $H_{F,\mathbf r}= H_F + V_{\mathbf r}$. This relation holds because the core-valence interaction $H_{c\text{-}v}$ is time-independent.
With that, one can further resolve
\begin{align}
	R_{n,\mathbf q} = 
	 \sum_{\mathbf r} {\mathcal M}_{\mathbf r}
	 e^{i\mathbf q\mathbf r} \sum_{n'} c_{\mathbf r,n'} [H_{F}+ V_{\mathbf r} - \Omega n' -\omega_o-\epsilon_{\rm c} -\epsilon_m^F-i\Gamma]^{-1} c^\dagger_{\mathbf r,n+n'},
\end{align}
where 
\begin{align}
	e^{iK(t)} c_{\mathbf r} e^{-iK(t)} &= \sum_n e^{in\Omega t} c_{\mathbf r,n},\\
	e^{iK(t)} c^\dagger_{\mathbf r} e^{-iK(t)} &= \sum_n e^{-in\Omega t} c^\dagger_{\mathbf r,n}
\end{align}
are the optically-dressed dipole transition operators [in analogy with the dressed RIXS operator defined in Eq.~\eqref{gesdfjagfds01}].

Altogether, this gives the final scattering matrix element for the Floquet Kramers-RIXS expression,
\begin{align}
	M_{m,j}(\mathbf q,\omega_i,\omega_o) &= 
	\sum_{\mathbf r} 
	 {\mathcal M}_{\mathbf r}
	 e^{i\mathbf q\mathbf r} \sum_{n,n'} \tilde s(\omega_l-\epsilon_m^F+\epsilon_j^F+n\Omega) \langle m^F| c_{\mathbf r,n'} \frac{1}{H_{F}+ V_{\mathbf r} - \Omega n' -\omega_o -\epsilon_c -\epsilon_m^F-i\Gamma} c^\dagger_{\mathbf r,n+n'} |j^F\rangle.
	\label{eq:M_final}
\end{align}
The structure of this matrix element is similar to the equilibrium one-- which is obtain from \eqref{eq:M_final} by fixing $n,n'=0$ and interpreting $|{m^F}\rangle, |j^F\rangle$ and $\epsilon_m^F, \epsilon_j^F$ as the eigenstates and energies of the time-independent Hamiltonian, respectively. Therefore, it allows for the following interpretation: 
Under periodic driving, and if the probe pulse $s(t)$ is long enough such that the factor $\tilde s(\omega)$ separates energies with difference $\Omega$ well, the equilibrium scattering cross-section is effectively replaced by a superposition of Floquet cross-sections, each of the latter comes with an individual Floquet component $c^{(\dagger)}_{\mathbf r,n}$ of the optically-dressed dipole transition operator. 

In general, these Fourier components are nonlocal. For example, within the high-frequency expansion, to leading order, the $n\neq 0$ component is proportional to $[c^{(\dagger)}_{\mathbf r}, H_n]$. For tight-binding systems, this amounts to hopping events occurring directly before or after the X-ray absorption/emission. The phenomenological interpretation of the Floquet Kramers-Heisenberg expression is that, upon X-ray absorption and emission, the electronic system can additionally absorb or emit Floquet photons from the optical drive. Thus, in the driven RIXS spectrum, additional absorption resonances in $\omega_i$ as well as additional loss and gain peaks with respect to $\omega_l$, both shifted by integer multiples of $\Omega$ from the static features, are generated.

\section{One-dimensional chain}

\subsection{Details of derivation}

As a simple example, suitable for numerical benchmarking, we consider a one-dimensional chain,
\begin{align}
H(t) = - J \sum_r (c_{r+1}^\dagger c_{r} e^{iA(t)} + h.c.).
\end{align}
In this section, we provide some details of the derivation of the resulting  expressions for the scattering operators, as well ad additional numerical results. 

Using the notation from above, there is just one site and one bond $\ell$ ($r-1\to r$)  per unit cell $r$, with $\delta_\ell=1$. 
The bond operators in momentum space are obtained as 
\begin{align}
B_{\pm}(q)
&= 
\frac{s_\pm}{2}
\sum_{k}
c_{k}^\dagger c_{k+q} 
\left(
e^{-i(k+ q)}
\pm
e^{ik}
\right).
\end{align}
To derive the RIXS operators we use $\mathcal J_{n,\ell}=\mathcal J_n(|A_0|) i^n$ and $\xi_\ell=0$ [cf.~Eq.~\eqref{genjsge}].
Then the form factors in Eq.~\eqref{hskaxz} become
\begin{align}
\mathcal R_{n,+}\, F_{q,+}\,
&=
-J\frac{s_n i }{z(z+in\Omega)}
(1+e^{iq})
\\
\mathcal R_{n,-}\, F_{q,-}\,
&=
-J
\frac{s_n (-1)^{n+1} }{z(z+in\Omega)}
(1-e^{iq})
\left(
1+\frac{2z}{in\Omega}
\right)
\end{align}
The RIXS operators  for even $n$ are
\begin{align}
R_{q,n}^F
&=
\mathcal J_{n} \cdot (
\mathcal R_{n,+}\, F_{q,+}
B_{+}(q)
+
\mathcal R_{n,-}\, F_{q,-}
B_{-}(q)
)
\\
&
= 
\frac{\mathcal J_{n} }{zn\Omega}
\sum_{k}
c_{k-q}^\dagger c_{k} 
\left(
\epsilon(k-q)-\epsilon(k)
\frac{z}{z+in\Omega}
\right),
\end{align}
with $\epsilon(k)=-2J\cos(k)$. For odd $n$,
\begin{align}
R_{q,n}^F
&=
\frac{\mathcal J_{n}  }{zn\Omega}
\sum_{k}
c_{k-q}^\dagger c_{k} 
\left(
iv(k-q)-iv(k)
\frac{z}{z+in\Omega}
\right),
\end{align}
with $v(k)=\partial_k\epsilon(k)=2J\sin(k)$.

\begin{figure}[t]
\centerline{\includegraphics[width=0.5\textwidth]{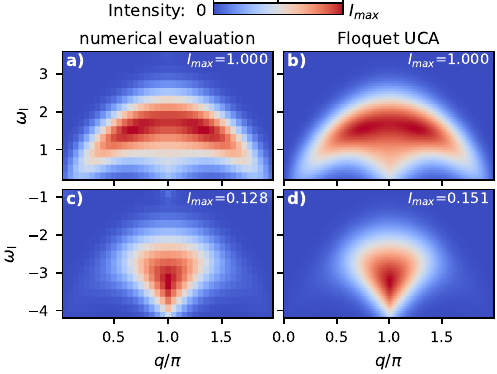}}
\caption{
Comparison of the analytical expression (Floquet UCA, right panels) and a numerical evaluation (left panels)  to the RIXS amplitude for the drive one-dimensional chain, analogous to the benchmarks presented in the main text, but for a shorter core-hole decay ($A_0=1.5$, $\Gamma=10$, $\Omega=4$, $\sigma=0.45$). }
\label{figS1}
\end{figure}

\subsection{Benchmark for larger $\Gamma/\Omega$}

For completeness, we present in Fig.~\ref{figS1} a benchmark similar to that in the main text, but for a shorter core-hole lifetime. Figure~\ref{figS1} shows the RIXS intensity for a drive with $A_0 = 1.5$ and $\Omega = 4$, and a core decay set by $\Gamma = 10$. The first and second rows of the figure show the signal in the spectral range of the main peak and the first Floquet sideband on the energy-gain side, respectively. As discussed in the main text, for a larger ratio $\Gamma/\Omega$, the RIXS intensity in the $n = 1$ Floquet sideband deviates from the current–current correlation function, in agreement between the numerical evaluation (Fig.~\ref{figS1}c) and the analytical expression (Fig.~\ref{figS1}d).

Finally, we remark that the data in Fig.~\ref{figS1} (as well as the data in the main text) have been computed in the presence of a core-hole potential $U_c = 2$. As anticipated from the analytical expression, the effect of $U_c$ on the Floquet sidebands is subleading in the expansions in $1/\Gamma$ and $1/\Omega$. Therefore, the numerical benchmark works well for both $U_c = 0$ and $U_c \neq 0$. The numerical evaluation of the full four-dimensional integral in Eq.~\eqref{app:rixs_formula_ME} is generally much more computationally costly when the core–valence interaction is included, because this complicates the evaluation of the correlation function $\langle P_{\mathbf r'}(t_1') P_{\mathbf r'}^\dagger(t_2') P_{\mathbf r}(t_2) P_{\mathbf r}^\dagger(t_1) \rangle$. We use a tensor-network-based method to compress this integrand and thereby accelerate the computations. This algorithm will be described elsewhere. (For the simple benchmark, a brute-force evaluation of the integrand is also possible, though numerically more expensive.)

\section{Symmetry decomposition of bond orders on the Kagome lattice}

\subsection{Decomposition of Bond operators}

\begin{table}[tbp]
\centering
\begin{tabular}{|c|c|c|c|c|c|c|c|c|c|c|}
\hline
 & $I$ & $T_{\mathbf a_i}$ & $C_2$ & $T_{\mathbf a_i}C_2$ & $C_3$ & $C_6$ & $\sigma_v$ & $T_{\mathbf a_i}\sigma_v$ & $\sigma_d$ & $T_{\mathbf a_i}\sigma_d$ \\
\hline
$|\mathcal{C}|$ & 1 & 3 & 1 & 3 & 8 & 8 & 6 & 6 & 6 & 6 \\
\hline
$A_1$ & 1 & 1 & 1 & 1 & 1 & 1 & 1 & 1 & 1 & 1 \\
$A_2$ & 1 & 1 & 1 & 1 & 1 & 1 & -1 & -1 & -1 & -1 \\
\hline
$B_1$ & 1 & 1 & -1 & -1 & 1 & -1 & 1 & 1 & -1 & -1 \\
$B_2$ & 1 & 1 & -1 & -1 & 1 & -1 & -1 & -1 & 1 & 1 \\
$E_1$ & 2 & 2 & -2 & -2 & -1 & 1 & 0 & 0 & 0 & 0 \\
$E_2$ & 2 & 2 & 2 & 2 & -1 & -1 & 0 & 0 & 0 & 0 \\
\hline
$F_1$ & 3 & -1 & 3 & -1 & 0 & 0 & 1 & -1 & 1 & -1 \\
$F_2$ & 3 & -1 & 3 & -1 & 0 & 0 & -1 & 1 & -1 & 1 \\
$F_3$ & 3 & -1 & -3 & 1 & 0 & 0 & 1 & -1 & -1 & 1 \\
$F_4$ & 3 & -1 & -3 & 1 & 0 & 0 & -1 & 1 & 1 & -1 \\
\hline
\end{tabular}
\caption{Character table of the group $C'''_{6v}$.}
\label{tabc6}
\end{table}

To classify the different bond orders, it is useful to group the bond operators 
$B_{\ell,\pm}=\sum_{\mathbf R} B_{\ell,\pm}(\mathbf R)$ according to symmetry \cite{Wagner2023}.
The relevant symmetry group is $C_{6v}'''$, which is generated by translations 
$T_{\mathbf a_i}$ of the lattice, $C_6$ rotations around the center of the hexagons, 
mirror reflections $\sigma_v$ through opposite bonds of the hexagons, and mirror 
reflections $\sigma_d$ through opposite sites of the hexagons. In addition, time-reversal 
symmetry changes the sign of the current operators $B_{\ell,-}$. 
The operators $B_{\ell,\pm}$ therefore form two representations 
$\mathcal D_{\pm}$ of this group. Using the character table shown in 
Table~\ref{tabc6}, the bond operators can be decomposed into basis sets which 
transform according to irreducible representations $D$ of the group.
The full bond representation decomposes as
\begin{align}
\mathcal D_{+} &= A_1 \oplus B_1 \oplus E_1 \oplus E_2 \oplus 2F_1 \oplus F_2 \oplus 2F_3 \oplus F_4,
\\
\mathcal D_{-} &= A_2' \oplus B_2' \oplus E_1' \oplus E_2' \oplus F_1' \oplus 2F_2' \oplus F_3' \oplus 2F_4',
\end{align}
where the prime indicates representations that are odd under time reversal.
This decomposition defines a symmetry-adapted bond basis
\begin{align}
B^{D}_j
=
\sum_{\ell} U^{D}_{j,\ell}
B_{\ell,\pm},
\label{eq:basisdef}
\end{align}
where $U^{D}_{j,\ell}$ ($j=1,\dots,d_D$) are orthonormal basis functions for the  irreducible representation $D$ with dimension $d_D$. Time-reversal even (odd) representations are formed from superpositions of 
$B_{\ell,+}$ ($B_{\ell,-}$). Within Landau theory, the expectation values 
$\langle B^{D}_j\rangle$ form $d_D$-dimensional order parameters.

\begin{figure}[t]
\centerline{\includegraphics[width=0.9\textwidth]{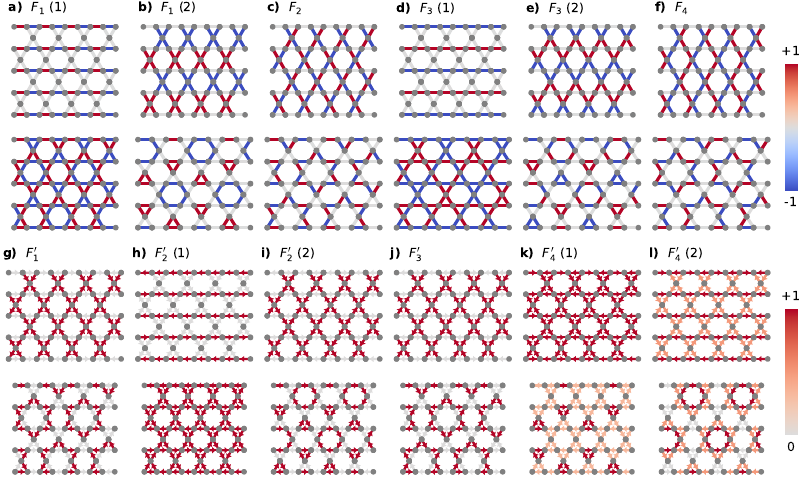}}
\caption{
Symmetry-adapted bond patterns $B^{D}_j$ for all three-dimensional irreducible 
representations $D$ of the real bond variables (a–f) and imaginary bond variables 
(g–l). For each panel, the upper plot shows the basis function $B^D_{1}$, 
corresponding to the $\mathbf M_1$ momentum component, by displaying the coefficients 
$U^D_{j,\ell}$ on all bonds $\ell$ (see color scale on the left). The lower plot shows the 
symmetric superposition $B^D_1+B^D_2+B^D_3$, corresponding to an equal 
superposition of the three ordering wavevectors. For representations with multiple 
copies ($F_1$, $F_3$, $F_2'$, $F_4'$), two distinct basis sets are shown. 
In the current patterns, the color of a bond indicates the magnitude of the 
current (see color bar), while the arrow indicates its direction (sign).
}\label{figS3}
\end{figure}

Of particular interest are the three-dimensional ($F$) representations, which break 
translational symmetry. The corresponding symmetry-adapted bond patterns are shown 
in Fig.~\ref{figS3}. The basis functions can be chosen such that the three components 
correspond to the three ordering wavevectors $\mathbf M_j$, i.e.
\begin{align}
T_{\mathbf a_i} B^{D}_j T_{\mathbf a_i}^{-1} 
= e^{i \mathbf M_j \cdot \mathbf a_i} B^{D}_j.
\label{eq:translation}
\end{align}
Moreover, for  some representations ($F_1$, $F_3$, $F_2'$, $F_4'$), two independent copies occur.  If possible, we single out one of the basis sets such that its
basis functions are localized on only one nearest-neighbor bond direction [see $F_1 (1)$, $F_3 (1)$, and $F_2' (1)$]. The two basis sets in $F_4'$ are instead constructed such that $F_4' (1)$ satisfies current conservation at the vertices and therefore represents an actual LCO.

\subsection{Floquet scattering form factors}

We now determine how these symmetry-breaking orders appear in the Floquet scattering 
signal. We restrict this analysis to the first sideband $n=1$. The analysis for higher sidebands 
can be performed analogously, but these sidebands are weaker and therefore experimentally 
harder to access. For $n=1$, the scattering operator 
\eqref{hskaxz}
becomes
\begin{align}
R^F_{\mathbf M_q,1}
=
\sum_{\ell}
\mathcal R_{n,+}\,
\mathcal J_{1,\ell}\,
F_{\mathbf M_q,\ell,+}\,
B_{\ell,-}
+
\sum_{\ell}
\mathcal R_{n,-}\,
\mathcal J_{1,\ell}\,
F_{\mathbf M_q,\ell,-}\,
B_{\ell,+}.
\end{align}
Using the inverse of Eq.~\eqref{eq:basisdef}, this can be expanded in the symmetry-adapted 
basis. With some  algebra,
\begin{align}
R^F_{\mathbf M_q,1}
=
\mathcal R_{n,+}
\sum_{D'\in F'}
F^{D'}_{\mathbf M_q}
B_{q}^{D'}
+
\mathcal R_{n,-}
\sum_{D\in F}
F^{D}_{\mathbf M_q}
B_{q}^{D},
\label{whaaszl23wa}
\end{align}
with form factors
\begin{align}
F^D_{\mathbf M_q}
&=
\sum_{\ell}
\mathcal J_{1,\ell}\,
F_{\mathbf M_q,\ell,+}\,
(U^{D}_{q,\ell})^*,
\\
F^{D'}_{\mathbf M_q}
&=
\sum_{\ell}
\mathcal J_{1,\ell}\,
F_{\mathbf M_q,\ell,-}\,
(U^{D'}_{q,\ell})^*.
\end{align}
Here  we have used that  choosing $\mathbf q=\mathbf M_q$ selects only the three-dimensional $F$ representations
 that break  translational symmetry, and, moreover, projects onto basis function $q$ of the respective representations.
The first and second term in \eqref{whaaszl23wa} show the overlap of the scattering operator with 
current type operators ($D'$ representations) and real bond operators ($D$ representations).
The form factors determine how the different order parameters 
$\langle B_{j}^{D}\rangle$ contribute to the scattering signal 
$\langle R^F_{\mathbf M_q,1}\rangle$. Because the Bessel factors 
$\mathcal J_{1,\ell}$ depend on the laser polarization $\mathbf A_0$, the 
scattering intensity exhibits characteristic polarization selection rules.

\subsection{Polarization selection rules}

\begin{figure}[t]
\centerline{\includegraphics[width=0.9\textwidth]{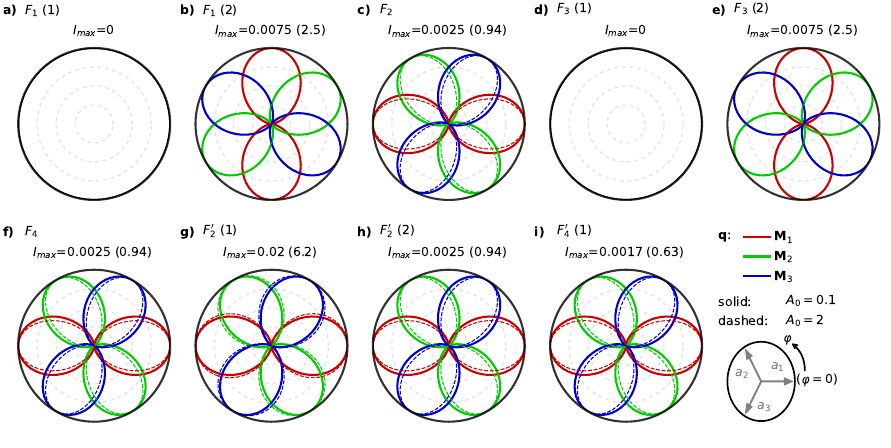}}
\caption{
Angular dependence of the Floquet form factors 
$F^D_{\mathbf M_q}(A_0,\varphi)$ for the three-dimensional irreducible 
representations $D$. Red, green, and blue curves correspond to the three 
ordering wavevectors $\mathbf M_1$, $\mathbf M_2$, and $\mathbf M_3$. Solid 
(dashed) lines show results for weak (strong) drive amplitude $A_0$. 
Representations $F_1$ and $F_3$ have maximal intensity when the laser 
polarization $\mathbf A_0$ is parallel to the ordering wavevector $\mathbf M_q$, 
whereas representations $F_2$ and $F_4$ are maximal when $\mathbf A_0$ is 
perpendicular to $\mathbf M_q$. Some representations, such as $F_1(1)$ and 
$F_3(1)$, give vanishing form factors due to geometric constraints, since the 
relevant bonds are perpendicular to $\mathbf M_q$. The relative weight of bond 
and current orders in the full scattering signal is determined separately by the 
prefactors $\mathcal R_{1,\pm}$ (see text).
}
\label{figS4}
\end{figure}

The angular dependence of the form factors 
$F^{D}_{\mathbf M_q}(A_0,\varphi)$ is shown in Fig.~\ref{figS4}. 
A clear symmetry-based selection rule emerges: representations $F_2$ and $F_4$ 
have maximal intensity for $\mathbf A_0 \perp \mathbf M_q$, whereas 
representations $F_1$ and $F_3$ are maximal for $\mathbf A_0 \parallel \mathbf M_q$. 
This follows from mirror symmetry with respect to a mirror plane parallel to 
$\mathbf M_q$. For $F_2$ and $F_4$, the $\mathbf M_q$ component is odd under this 
mirror, while for $\mathbf A_0 \parallel \mathbf M_q$ the vector potential is even, forcing the matrix element to vanish  in this case. The opposite parity applies to $F_1$ and 
$F_3$, for which the signal vanishes when $\mathbf A_0 \perp \mathbf M_q$. 

In addition, some representations have vanishing intensity due to geometric 
constraints. In particular, for the $F_1(1)$ and $F_3(1)$ patterns all bonds are 
perpendicular to $\mathbf M_q$, which causes the Bessel factors to vanish for the 
allowed polarization direction. As a result, these orders produce no signal in the 
first Floquet sideband. Finally, real bond orders and current orders are distinguished by the prefactors 
$\mathcal R_{1,\pm}$, which depend on the ratio between the intermediate-state 
detuning $|z|$ and the drive frequency $\Omega$. In the limit $|z| \gg \Omega$, 
one finds $|\mathcal R_{n,-}| \gg |\mathcal R_{n,+}|$,
so that real bond orders dominate the signal.
When the two energy scales are comparable, current orders can dominate, 
with the $F_2'$ loop-current pattern producing the largest signal. In systems where 
multiple orders coexist, the polarization direction of maximum intensity is therefore 
expected to rotate as a function of the ratio $\Omega/|z|$, which could be used as an additional
fingerprint. 

\end{appendix}

\end{document}